\documentclass[10pt,twocolumn,twoside,submit]{JCNtran}
\usepackage{multirow}
\usepackage{amsmath}
\usepackage{color}
\usepackage{amsxtra}
\usepackage{amssymb}
\usepackage{threeparttable}
\usepackage{graphicx}
\usepackage{verbatim}   % useful for program listings
\usepackage{subfigure}  % use for side-by-side figures
\usepackage{makeidx}  % allows for index generation
\usepackage[dvips]{epsfig}
\usepackage[latin1]{inputenc}
\usepackage[T1]{fontenc}

\def\BibTeX{{\rm B\kern-.05em{\sc i\kern-.025em b}\kern-.08em
    T\kern-.1667em\lower.7ex\hbox{E}\kern-.125emX}}

\begin{document}
\bibliographystyle{jcn}

\title{Fixed-complexity Sphere Encoder \\ for Multi-user MIMO Systems}
\author{Manar Mohaisen and KyungHi Chang
\thanks{Manuscript received March 22, 2009; approved for publication by (), Division III Editor, June 11, 2009.}
\thanks{This work was supported by Inha University Research Grant.}
\thanks{M. Mohaisen and K.H. Chang are with the Graduate School of Information Technology and Telecommunications, Inha University, Incheon, Korea, emails:
manar.subhi@kut.ac.kr, khchang@inha.ac.kr.}} \markboth{JOURNAL OF
COMMUNICATIONS AND NETWORKS, VOL. 5, NO. 3, SEPTEMBER 2009}{Mohaisen
\lowercase{\textit{et al}}.: Fixed-complexity Sphere Encoder ...}
\maketitle

\begin{abstract}
In this paper, we propose a fixed-complexity sphere encoder (FSE)
for multi-user MIMO (MU-MIMO) systems. The proposed FSE accomplishes
a scalable tradeoff between performance and complexity. Also,
because it has a parallel tree-search structure, the proposed
encoder can be easily pipelined, leading to a tremendous reduction
in the precoding latency. The complexity of the proposed encoder is
also analyzed, and we propose two techniques that reduce it.
Simulation and analytical results demonstrate that in a $4 \times 4$
MU-MIMO system, the proposed FSE requires only 11.5$\%$ of the
computational complexity needed by the conventional QRD-M encoder
(QRDM-E). Also, the encoding throughput of the proposed encoder is
7.5 times that of the QRDM-E with tolerable degradation in the BER
performance, while achieving the optimum diversity order.
\end{abstract}


\vspace{10pt}
\section{\uppercase{Introduction}}
\label{sec:introd}
Multi-user multiple-input multiple-output (MU-MIMO) techniques are becoming increasingly important as the base station (BS) should have the capability to  simultaneously communicate with a large number of users. To this end, dirty paper coding (DPC), which is shown to achieve the capacity region of the Gaussian MIMO broadcast channel \cite{Shamai}, has been proposed by Costa \cite{Costa}.\\
\indent Several MU-MIMO precoding schemes were proposed in the literatures in order to achieve the near-capacity. Linear zero-forcing (ZF) precoding was introduced in \cite{Peel}, where the transmitted vector is pre-filtered using the pseudo-inverse of the channel matrix. As a consequence, a high transmission power is required, particularly when the channel matrix is ill-conditioned. To overcome this problem, regularized channel inversion, i.e., linear minimum mean square error (MMSE) precoder, was proposed to reduce the required transmission power of the ZF precoding scheme, while achieving a tradeoff between interference and noise amplification \cite{Peel}. Moreover, Tomlinson-Harashima precoding (THP) scheme achieves better performance by limiting the transmit power, via a non-linear modulo operation \cite{Tomlinson}, \cite{Harashima}. Also, it is shown that the coding loss due to the modulo operation vanishes for high-order modulation schemes. Transmit power can be further reduced by perturbing the transmitted vector, as in \cite{Hochwald}, where the optimum perturbation vector is found using the sphere encoder (SE). Although SE has a small average computational complexity, its worst-case complexity is very high. Besides, SE is sequential in the tree-search phase, which limits the potential for efficient hardware implementation. To overcome the random complexity of the SE, the QR-decomposition with M-algorithm encoder (QRDM-E) was proposed in \cite{QRDME}. The main idea of the QRDM-E is to retain a fixed number of candidates at each encoding level. Although QRDM-E achieves the same performance of the SE, its complexity is high and its tree search strategy limits the possibilities for the efficient hardware implementation using pipelining.\\
\indent In this paper, we propose a fixed-complexity sphere encoder (FSE) based on the fixed-complexity sphere decoder \cite{{Barbero1}}, that achieves the optimum diversity order of the QRDM-E, as well as a flexible tradeoff between performance and complexity. Moreover, because the proposed FSE has a parallel search structure, it can be pipelined, which tremendously reduces the precoding latency. In addition, we evaluate the complexity of the conventional schemes and the proposed FSE and introduce two techniques that reduce its complexity.\\
\indent The rest of this paper is organized as follows. In Section II, we introduce the system model and the statement of problem. In Section III, we review the conventional vector-perturbation techniques. In Section IV, we introduce the proposed FSE and in Section V we analyze its complexity and those of the SE and QRDM-E. Two proposed methods to reduce the complexity of the FSE are introduced in Section VI and simulation results are shown in Section VII. Finally, conclusions are drawn in Section VIII.
%%%%%%%%%%%%%%%%%%%%%%%%%%%%%%%%%%%%%%%%%%%
\vspace{10pt}
\section{\uppercase{System Model and the Problem Statement}}
\label{sec:SysmModel}
We consider a downlink MU-MIMO transmission system in which a BS with $N_t$ transmit antennas communicates simultaneously with $N_u$ decentralized single-antenna users. Without loss of generality, we assume that $N = N_t = N_u$. Also, we consider a flat-fading and slowly time-varying channel, whose state information is perfectly known at the transmitter, if not otherwise mentioned. Then, the system is converted to the $K$-dimensional real lattice problem, where $K = 2N$.\\
\indent Let $\textbf{H} \in {\mathbb{R}}^{K\times K}$ denote the channel matrix, and  $\textbf{s} \in {\mathbb{R}}^K$ denote the data vector. \\
\indent Linear precoding techniques are the simplest where in the
case of linear zero-forcing precoding (LZF) the effect of the
channel is canceled by precoding the transmitted data vector using
the pseudo-inverse of the channel matrix.
\begin{equation}
\textbf{x}_{\text{zf}} = \frac{1}{\sqrt{\gamma}} \textbf{H}^{\dagger} \textbf{s},
\end{equation}
where the scaling factor ${\gamma}$ is present to fix the \textit{expected} total transmit power to ($P_T$); that is,
\begin{equation}
\gamma = \frac{1}{\text{P}_T} \mathrm{Tr}\left\{\left(\textbf{HH}^H\right)^{-1}\right\},
\end{equation}
where $\mathrm{Tr}(\cdot)$ refer to the trace operation. As a consequence, the receive SNR at any MS is given by:
\begin{eqnarray}
 \mbox{SNR} = \frac{\mbox{E}(ss^*)}{\gamma \sigma_n^2}.
\end{eqnarray}
If the channel matrix is ill-conditioned, $\gamma$ becomes large and consequently the post-processing signal to noise ratio (SNR) is decreased. To partially overcome this drawback, linear minimum mean-square error (MMSE) precoding  can be used to regularize the channel matrix. The precoded signal using LMMSE is therefore given by:
\begin{equation}
\textbf{x}_{\text{mmse}} = \textbf{H}^H\left(\textbf{HH}^H + \alpha \textbf{I}_{K}\right)^{-1}\textbf{s},
\end{equation}
where $\alpha$ = $K\sigma_n^2/P_T$ is the regularization factor. Although the LMMSE precoder reduces the required transmit power, i.e., reduces $\gamma$, its performance is still mediocre and further improvement can be therefore achieved.\\
\indent Unlike the LMMSE precoder which regularizes the channel matrix, the non-linear THP algorithm works on the data vector \textbf{s} so that the required transmit power is reduced \cite{Tomlinson}, \cite{Harashima}. Hence, a linear representation of the THP algorithm can be seen as finding the perturbed vector
%\indent The main goal of the vector-perturbation is to create the vector
\begin{equation}
\tilde{\textbf{s}} = \textbf{s} + \tau \textbf{t},
\label{eq:pv}
\end{equation}
such that the required transmit power of the precoded vector is reduced. In (\ref{eq:pv}), $\tau$ is an integer that depends on the employed modulation scheme, and \textbf{t} is a \textit{K}-dimensional integer vector. In \cite{Hochwald}, $\tau$ is given by:
\begin{equation}
\tau = 2\left(|c|_{max} + \Delta/2\right),
\end{equation}
where $|c|_{max}$ is the absolute value of the constellation point with the largest magnitude, and $\Delta$ is the spacing between the constellation points. Note that THP algorithm finds the elements of \textbf{t} in a successive way, where the \textit{t} candidate that minimizes the required transmit power at each encoding level is retained. This is equivalent to the successive interference cancellation in the signal detection literature.\\
\indent Although THP algorithm reduces the required transmit power
compared to the linear precoding schemes, better performance can be
obtained by optimally perturbing the transmit vector so that further
reduction in the transmit power is obtained. The vector perturbation
can be represented as an integer-lattice search, where at the
transmitter, \textbf{t} is chosen such that $\gamma$ is minimized;
that is,
\begin{align}
\textbf{t} &= \underset{\textbf{t} \in {\mathbb{Z}}^K}{\operatorname{arg\,min}} \, \left\{(\textbf{s} + \tau \textbf{t})^T \textbf{P}^T\textbf{P}(\textbf{s} + \tau \textbf{t})\right\},\nonumber\\
          &= \underset{\textbf{t} \in {\mathbb{Z}}^K}{\operatorname{arg\,min}} \, \left\|\textbf{P}(\textbf{s} + \tau \textbf{t})\right\|^2.
\label{eq:FSE}
\end{align}
Let the transpose of the matrix $\textbf{H}$ be factorized into the product of a unitary matrix \textbf{Q} and an upper triangular matrix \textbf{R}, thus, the search problem in (\ref{eq:FSE}) based on the zero-forcing criterion is simplified to:
\begin{align}
\textbf{t} &= \underset{\textbf{t} \in {\mathbb{Z}}^K}{\operatorname{arg\,min}} \, \left\|\textbf{L}(\textbf{s} + \tau \textbf{t})\right\|^2,\nonumber\\
                     &= \underset{\textbf{t} \in {\mathbb{Z}}^K}{\operatorname{arg\,min}} \, \sum_{i = 1}^{K} \left\|L_{i, i}(s_i + \tau t_k) + \sum_{j=1}^{i-1} L_{i, j}(s_j + \tau \hat{t}_j)\right\|^2,
\label{eq:FSE1}
\end{align}
where the lower triangular matrix $\textbf{L}$ equals $\left(\textbf{R}^{-1}\right)^T$. When the MMSE criterion is used, the extended matrix $\tilde{\textbf{H}} = [\textbf{H}^T\,\,\sqrt{\alpha}\textbf{I}]^T$ is factorized into the \textbf{Q} and \textbf{R} matrices, where $\textbf{L}$ also equals $\left(\textbf{R}^{-1}\right)^T$. Due to the QR-decomposition property \cite{Wubben}
\begin{equation}
\left[
    \begin{array}{c} \textbf{H}^T \\  \sqrt{\alpha}\textbf{I}\end{array}
\right] = \left[
    \begin{array}{c} \textbf{Q}_1 \\  \textbf{Q}_2 \end{array}
\right] \textbf{R} = \left[
    \begin{array}{c} \textbf{Q}_1\textbf{R} \\
    \textbf{Q}_2\textbf{R} \end{array}
\right],
\end{equation}
it holds that $\textbf{R}^{-1} = \textbf{Q}_2/\sqrt{\alpha}$ \cite{Wubben}. By definition $\sqrt{\alpha}$ is a strictly positive real number, then it does not affect the search result in (\ref{eq:FSE1}). Therefore, $\textbf{L} = \textbf{Q}_2^T$ also leads to the required perturbation without the need for explicitly inverting $\textbf{R}$.\\
\indent In this paper, $t_k$ is selected from the symmetric integer set:
\begin{equation}
{\mathcal{A}} = [-a,\,-a + 1, \cdots,\,a -1,\,a],
\end{equation}
where $a$ is a positive integer chosen to achieve a tradeoff between performance  and complexity of the vector-perturbation algorithm. Therefore, as $a$ increases, the bit error rate (BER) performance is improved but the complexity is also increased, and vice-versa. Hereafter, $T = (2a + 1)$ denotes the number of elements in the set $\mathcal{A}$. Note that $a$ will be optimized using extensive simulation.\\
%%%%%%%%%%%%%%%%%%%%%%%%%%%%%%%%%%%%%%
\vspace{10pt}
\section{\uppercase{Review of the Conventional Vector-Perturbation Techniques}}
\subsection{Sphere Encoder}
The idea of the SE is to limit the search to the vectors \textbf{t} which resides in a hypersphere with a predefined radius. Therefore,
\begin{equation}
\textbf{t}_{\text{SD}} = \underset{\textbf{t} \in {\mathbb{Z}}^K}{\operatorname{arg\,min}} \, \left(\left\|\textbf{P}(\textbf{s} + \tau \textbf{t})\right\|^2 \leq d^2\right),
\end{equation}
where $d$ is the predefined search radius. The search radius is updated when a vector \textbf{t} with smaller accumulative metric is found.\\
\indent The main drawbacks of the SE is that it has a random complexity that can be inapplicable at the worst-case. Also, SE has a sequential tree-search phase which limits the efficient hardware implementation.
%%%%%%%%%%%%%%%%%%%%%%%%%
%%%%%%%%%%%%%%%%
\subsection{QR-decomposition with M-algorithm Encoder (QRDM-E)}
In QRDM-E, the best $M$ branches that have the least accumulative metrics are retained at each encoding level. The value of $M$ is used to set a tradeoff between performance and computational complexity, where to accomplish a fair comparison with the FSE, $M$ is set to $T$. Therefore, at the first tree-search stage, the best $M$ branches are retained for level 2. At level 2, the retained branches are expanded to all possible combinations of ($s_2 + \tau t_k$). The resulting $M^2$ branches are sorted according to their accumulative metrics calculated via (\ref{eq:FSE1}), where only the $M$ branches with the smallest accumulative metrics are retained for level 3. This strategy is repeated up to the last encoding level, where the perturbed vector $\tilde{\textbf{\textit{s}}}$ that has the smallest accumulative metric is precoded and transmitted.\\
\indent The QRDM-E algorithm has a fixed complexity that is independent of the noise variance or the channel conditioning. Nonetheless, its complexity highly increases for high $K$ and $T$.
%%%%%%%%%%%%%%%%%%%%%%%%%%%%%%
\begin{figure*}[!t]
\centering
\includegraphics[width=11.5cm]{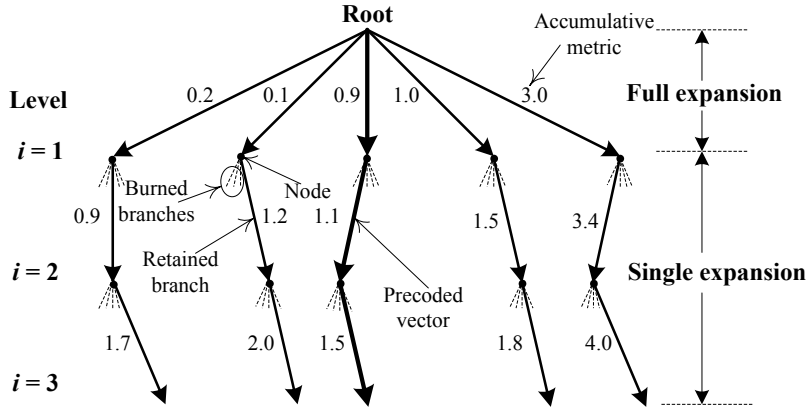}
\caption{Example of the proposed fixed-complexity sphere encoder for
$K = 3$ and $T = 5$.} \label{FSE}
\end{figure*}
%%%%%%%%%%%%%%%%%%%%%
\vspace{10pt}
\section{\uppercase{Proposed Fixed-Complexity Sphere Encoder}}
The proposal of the FSE is motivated by:
\begin{itemize}
    \item \textbf{Increasing the encoding throughput:} The encoding throughput of the QRDM-E is fixed which is favorable for communication systems. Nevertheless, this encoding throughput is low due to the limited possibilities for efficient hardware implementation of the QRDM-E. This is mainly due to the search strategy employed at the QRDM-E where high number of metrics are compared at each encoding level. Our proposed FSE has a high encoding throughput due to its parallel tree search phase.
    \item \textbf{Decreasing the complexity:} The complexity of the QRDM-E significantly increases for high $K$ and $T$. Also, the SE has a high worst-case computational complexity.
\end{itemize}
\indent To overcome these drawbacks of the conventional encoding schemes, the FSE is proposed in this paper.\\
\indent The tree-search phase of the proposed FSE algorithm consists
of the following two steps:
\begin{itemize}
    \item \textbf{Full expansion:} At the first $p$ tree search levels, the retained branches are expanded to all possible nodes, and all the resulting branches are retained for the next level. The choice of $p$ is addressed in Section VII.
    \item \textbf{Single expansion:} Only a single expansion is performed from each retained nodes at the precedent encoding level. This is done by following the decision-feedback equalization (DFE) path.
\end{itemize}
At the last search level, the metrics of the obtained perturbed vectors $\tilde{\textbf{s}}_1,\,\tilde{\textbf{s}}_2, \cdots, \tilde{\textbf{s}}_T$ are compared, and the vector that has the smallest metric is precoded and transmitted.\\
%%%%%%%%%%%%%%%%%%%%%
\begin{figure}[!t]
\centering
\includegraphics[width=8.4cm]{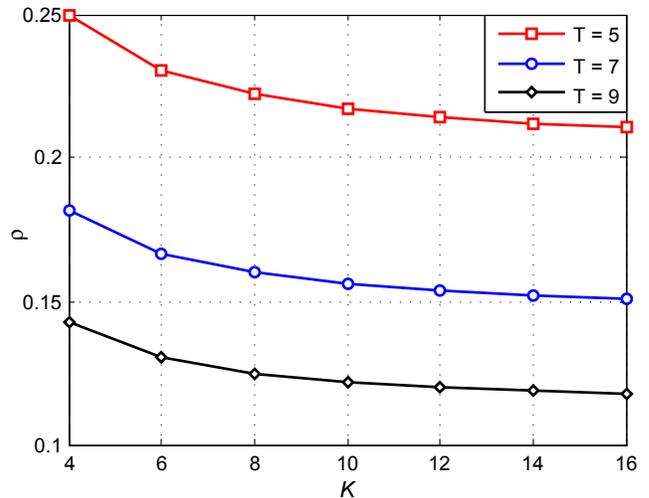}
\caption{Ratio between the complexities of FSE and QRDM-E for
several $K$ and $T$ values.} \label{fig:Rho}
\end{figure}
%%%%%%%%%%%%%%%%%
\indent Fig. \ref{FSE} shows an example of the proposed FSE
algorithm for $p = 1$, $K = 3$, and $t_k \in
\left\{-2,\,-1,\,0,\,1,\,2\right\}$. At the first search level, the
root node is extended to the five possible combinations of $(s_1 +
\tau t_k), \text{ where } k\,=\,1,\,2,\,\cdots,\,5$. The metrics of
the resulting branches are then calculated, and all the branches are
retained for the next level. At levels $2$ to $K$, only a single
expansion is performed from each retained node. This is done by
following the decision-feedback equalization path. At the last
search level, the vector $\tilde{\textbf{s}}$ that has the lowest
accumulative metric, indicated by the thick line in Fig. \ref{FSE},
is precoded and transmitted.
%%%%%%%%%%%%%%%%%%%%%%%%%%%
\vspace{10pt}
\section{\uppercase{Analysis on the Computational Complexity}}
In this section, we compute the computational complexity of the vector-perturbation techniques in terms of the number of visited nodes, i.e., number of metric computations.\\
\indent The worst-case complexity of the SE is given by:
\begin{align}
\text{C}_{\text{SE}} &= \sum_{i=1}^K T^i,\nonumber\\
                                         &= \frac{T^{K+1} - T}{T - 1}.
\end{align}
For high $K$, the worst-case complexity of the SE becomes high and inapplicable.\\
\indent To obtain the perturbation vector, the QRDM-E algorithm performs
\begin{equation}
\text{C}_{\text{QRDM-E}} = T + (K-1)T^2
\end{equation}
metric computations. On the other hand, the proposed FSE only requires
\begin{equation}
\text{C}_{\text{FSE}} = KT
\end{equation}
metric computations. This demonstrates that the proposed algorithm is light and suitable for mobile communication systems which are power and latency limited.\\
\indent Figure \ref{fig:Rho} shows the ratio $\rho = \text{C}_{\text{FSE}} / \text{C}_{\text{QRDM-E}}$ versus the real space dimension $(K = 2 \times N_t)$. In a $4 \times 4$ system, i.e., $K = 8$, and $T = 7$, FSE requires only 16$\%$ of the computational complexity of the QRDM-E.\\
\indent In \cite{Barbero2}, it has been shown that at the same clock frequency of 100 MHz, $4\times 4$ system, and 16-QAM, the achieved throughput of the FSE and the QRDM-E are 400 Mbps and 53.3 Mbps, respectively.\footnote{In fact, these results are given for the hardware implementation of the FSD and QRD-M detection algorithms. Since, the tree search phase is similar for the encoding and the detection, we consider these values appropriate for the purpose of comparison.} This shows that the encoding throughput of the FSE is 7.5 times that of the conventional QRDM-E.
%%%%%%%%%%%%%%%%%%%%%%%%%%%%%%%%%%%%%%%%%%%
\vspace{10pt}
\section{\uppercase{Proposed Complexity Reduction Techniques for FSE}}
\indent To reduce the computational complexity of the FSE algorithm,
we propose utilization of pre-computations to obtain the elements of
$\textbf{L}(\textbf{s} + \tau \textbf{t})$, which are accessed via
their indices. These elements are saved in the matrix $\textbf{A}
\in {\mathbb{R}}^{U \times D \times T}$, where $U = (\sum_{i =
1}^N\,\,i)$ and \textit{D} is the size of the real constellation
set. As these computations are only performed each time the channel
matrix is updated, the number of multiplication and addition
operations required for each transmission are given by:
\begin{equation}
\text{C}_{\text{p}}^{\text{mul}} = \frac{DTK(K+1) + 2T - 2}{2N_f}
\end{equation}
and
\begin{equation}
\text{C}_{\text{p}}^{\text{add}} = \frac{(T-1)D}{N_f},
\end{equation}
where $\text{C}_{\text{p}}^{\text{mul}}$ and $\text{C}_{\text{p}}^{\text{add}}$ are the required real multiplication and addition operations, respectively, for the pre-computation stage. Also, $N_f$ is the number of transmissions using the same channel state information (CSI).\\
\indent At the tree-search phase, instead of comparing the second norms of the branch metrics, as in (\ref{eq:FSE1}), we propose to compare the absolute values of the branch metrics, hereafter referred to as the absolute metric. This is done by comparing the obtained metrics in (\ref{eq:FSE1}), before the square operation. Therefore, the branch with the smallest absolute metric is selected, and its accumulative metric is computed, as in (\ref{eq:FSE1}). As a consequence, the number of required multiplication operations at each node is reduced from $T$ to one. In the sequel, this technique is referred to as \textit{comparison-before-squaring strategy}. Hence, the multiplication and addition operations required at the tree search phase of the proposed FSE for algorithm ($p = 1$) are as follows:
\begin{equation}
\text{C}_{\text{ts}}^{\text{mul}} = KT
\end{equation}
and
\begin{equation}
\text{C}_{\text{ts}}^{\text{add}} = \frac{1}{2} T^2 K (K-1) + T - 1.
\end{equation}
Then, the total number of multiplication operations required by the proposed FSE algorithm is given by:
\begin{equation}
\text{C}^{\text{mul}} = \frac{DTK(K+1) + 2T - 2}{2N_f} + KT,
\end{equation}
and the total number of additions is given by:
\begin{equation}
\text{C}^{\text{add}} = \frac{D(T-1)}{N_f} + \frac{1}{2} T^2 K (K-1) + T -1.
\end{equation}
For large $N_f$, $\text{C}^{\text{mul}} \approx \text{C}_{\text{ts}}^{\text{mul}}$ and $\text{C}^{\text{add}} \approx \text{C}_{\text{ts}}^{\text{add}}$.
%%%%%%%%%%%%%%%%%%%%%%%%%%%%%%%%%%%%%%%%%%%%%%%%
\begin{figure}[t]
\centering
\includegraphics[width=8.4cm]{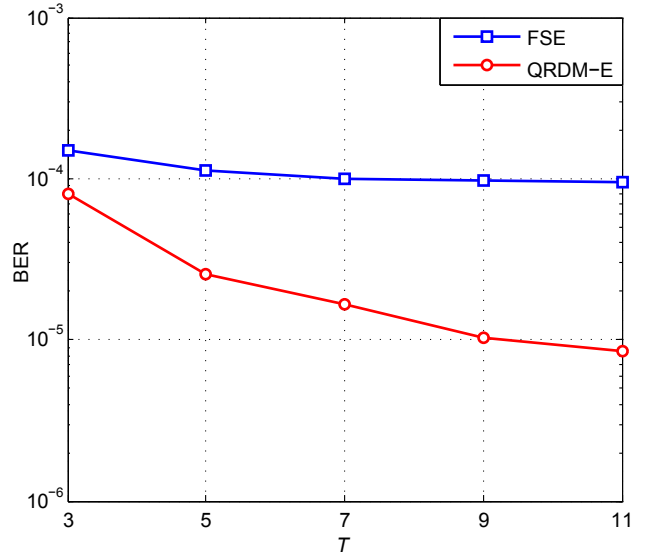}
\caption{BER performance of the vector perturbation techniques for $K = 8$, several values of $T$, using QPSK modulation, and at SNR of 20dB.}
\label{fig:Toptimize}
\end{figure}
%%%%%%%%%%%%%%%%%%%
%\begin{figure}[t]
%\centering
%\includegraphics[width=8.4cm]{BER}
%\caption{BER performance of the vector perturbation techniques for $K = 8$, $T = 7$, and using QPSK modulation.}
%\label{fig:BER}
%\end{figure}
%%%%%%%%%%%%%%%%%%
\begin{figure}[t]
\centering
\includegraphics[width=8.4cm]{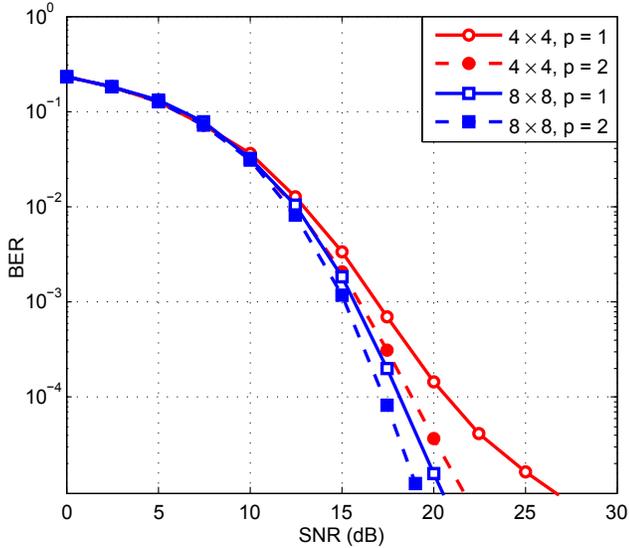}
\caption{BER performance of the proposed fixed-complexity sphere encoder for $T = 3$, $p = 1$ and $p = 2$ in a $4 \times 4$ and $8 \times 8$ systems, using QPSK modulation.}
\label{fig:p1p2}
\end{figure}
%%%%%%%%%%%%%%%%%%%%%%%%%%%%%%%%%%%
\vspace{10pt}
\section{\uppercase{Simulation Results and Discussion}}
\label{sec:Simulations}
\subsection{\uppercase{Simulations with Perfect CSI at the Transmitter}}
\label{sec:Pcsi}
In this section, we investigate the bit error rate (BER) performance of the conventional THP scheme, QRDM-E, and the proposed FSE in $4\,\times\,4$ and $8\,\times\,8$ MU-MIMO systems, i.e., for $K = 8$ and $K = 16$, respectively, using QPSK modulation. The conventional THP algorithm is considered as the special case of the proposed FSE algorithm when the branch with the smallest accumulative metric is the only one retained at each precoding level, i.e., if only the decision-feedback equalization path is followed \cite{Liu}. In the sequel, the MMSE precoding criterion is used due to its superior performance compared with the ZF criterion.\\
\indent Figure \ref{fig:Toptimize} shows the BER of the FSE and QRDM-E schemes for several sizes of the set $\mathcal{A}$ at SNR = 20 dB. We remark that the best improvement in the performance happens when moving from $T = 3$ to $T = 5$. For $T \geq 9$, no further performance improvement is remarked in the case of the FSE while a small additional improvement is remarked in the case of the QRDM-E. Therefore, as a tradeoff between performance and complexity, we set $T$ to $9$ in the sequel.\\
%\indent Figure \ref{fig:BER} shows the BER performance of the introduced vector perturbation techniques with the MMSE precoding for $K = 8$, using QPSK modulation. At a target BER of $10^{-4}$, the proposed FSE algorithm outperforms the THP by 6dB and lags the performance of the QRDM-E by 2dB. We notice also that the proposed FSE achieved the full diversity order of the QRDM-E, i.e., their BER curves are parallel. This degradation in the BER performance is tolerable as compared to the huge gain in the encoding throughput and the reduction in the computational complexity.
\indent Figure \ref{fig:p1p2} depicts the BER performance of the proposed FSE for $p = 1$ and $p = 2$, referred to as FSE-$p1$ and FSE-$p2$ in the sequel, in $4 \times 4$ and $8 \times 8$ systems. In $4 \times 4$ system and for $T = 3$, FSE-$p1$ and FSE-$p2$ require 24 and 66 metric computations, respectively. At target BER of $10^{-4}$, FSE-$p2$ outperforms FSE-$p1$ by 2 and 0.9 dB in $4 \times 4$ and $8 \times 8$ systems, respectively. Note that the FSD \cite{Barbero1} does not enjoy better performance when the complexity is increased at $4 \times 4$ system, while as it is aforementioned, the FSE has better performance when moving from FSE-$p1$ to FSE-$p2$.\\
%%%%%%%%%%%%%%%%%%%%%%%%%%%
%%%%%%%%%%%%%%%%%%%%%%%
\begin{figure}[t]
\centering
\includegraphics[width=8.4cm]{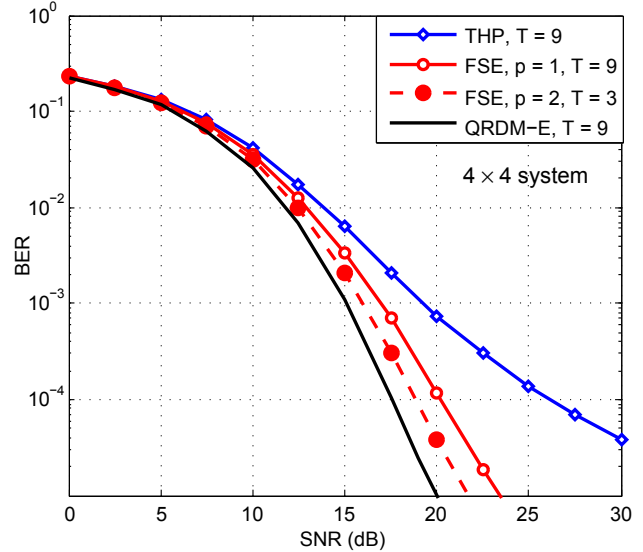}
\caption{BER performance of the proposed fixed-complexity sphere encoder in a $4 \times 4$ system, and using QPSK modulation.}
\label{fig:ber44}
\end{figure}
%%%%%%%%%%%%%%%%%%%%%%%%%%%%%
%%%%%%%%%%%%%%%%%%%%%%
\begin{figure}[t]
\centering
\includegraphics[width=8.4cm]{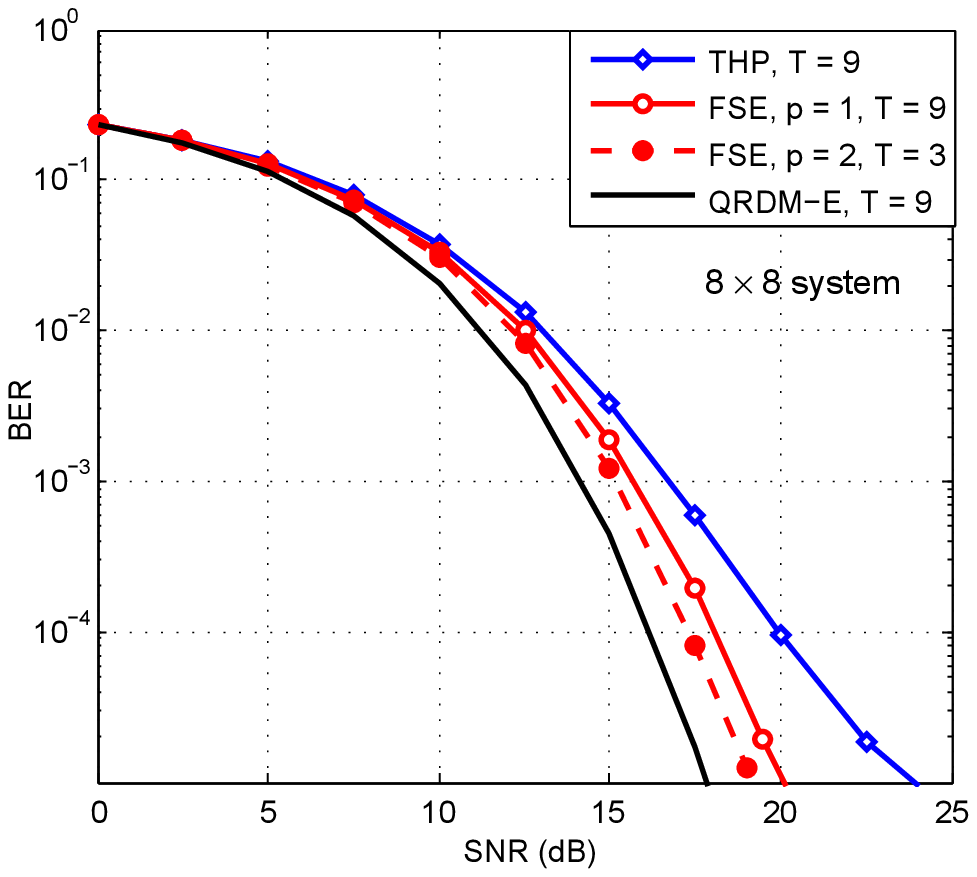}
\caption{BER performance of the proposed fixed-complexity sphere encoder in a $8 \times 8$ system, and using QPSK modulation.}
\label{fig:ber88}
\end{figure}
%%%%%%%%%%%%%%%%%%%%%%%%%%%
\begin{table}
  \caption{Computational complexity of the vector-perturbation schemes in terms of the number of visited nodes.}
  \begin{tabular}{| l | c | c | c |}
  \hline
        System & QRDM-E & FSE-$p1$ ($T = 9$)    & FSE-$p2$ ($T = 3$) \\
  \hline
  $4 \times 4$ & 576    &  72                  & 66                 \\
  \hline
  $8 \times 8$ & 1224   &  144                 & 138                \\
  \hline
  \end{tabular}
  \label{tbl:cx}
\end{table}
%%%%%%%%%%%%%%%%%%%%%%%%%%%%%%%%%%%%
%%%%%%%%%%%%%%%%%%%%%%%%%%%%%%%%%%%%%%%%
\begin{table}
  \caption{Mean and standard deviation of the metrics corresponding to the retained candidates at the last encoding level averaged over 100,000 independent channel realizations.}
  \centering
  \begin{tabular}{| l | c | c | c | c |}
  \hline
        \multirow{2}{*}{System} & \multicolumn{2}{|c|}{FSE-$p1$ ($T = 9)$} & \multicolumn{2}{|c|}{FSE-$p2$ ($T = 3)$} \\
  \cline{2-5}
  & mean & std & mean & std \\
  \hline
  $4 \times 4$ & 40.7 & 40.4 & 16.7 & 13.8 \\
  \hline
  $8 \times 8$ & 22.2 & 15.2 & 10.7 & 3.9 \\
  \hline
  \end{tabular}
  \label{tbl:var}
\end{table}
%%%%%%%%%%%%%%%%%%%%%%%%%%%%%%%%%%%%%%%5
\indent Figures \ref{fig:ber44} and \ref{fig:ber88} depict the BER performance of the proposed FSE compared to those of the conventional algorithms in $4 \times 4$ and $8 \times 8$ systems, respectively. Table \ref{tbl:cx} gives the computational complexities of the vector perturbation algorithms. The stated results demonstrate the light complexity of the proposed FSE compared to the conventional QRDM-E. For instance, FSE-$p1$ and FSE-$p2$ require only 12.5$\%$ and 11.46$\%$ of the number of metric computations performed by the conventional QRDM-E. In terms of BER performance, in $4 \times 4$ system, FSE-$p2$ outperforms THP technique by 7.4dB while lagging the performance of the QRDM-E by 1.3dB at a target BER of $10^{-4}$. In $8 \times 8$ system, FSE-$p2$ outperforms THP technique by 2.7dB while lagging the performance of QRDM-E by 1.2dB at the same target BER.\\
\indent From Figures \ref{fig:ber44} and \ref{fig:ber88}, it is
clear that FSE-$p2$ outperforms FSE-$p1$ although it has lower
computational complexity. This is because the candidates retained by
the FSE-$p2$ are much closer to each other in the Euclidean space.
As a consequence, the convergence to the optimum candidate is more
probable than in the case of FSE-$p1$ which leads to more distant
candidates in the Euclidean space. Table \ref{tbl:var} depicts the
mean and standard deviation of the accumulative metrics
corresponding to the retained vector candidates at the last encoding
level. It is evident that FSE-$p2$ leads to both low mean and
standard deviation of those metrics as compared to the FSE-$p1$.
Hence, we recommend using FSE-$p2$ instead of FSE-$p1$ even in
low-dimensional MU-MIMO systems.
%\begin{figure}[t]
%\begin{center}
%\epsfxsize=8cm \leavevmode\epsfbox{./solu.eps} \caption{Network
%management by using a mobile agent.} \label{fig:sol_am}
%\end{center}
%\end{figure}
%%%%%%%%%%%%%%%%%%%%%%%%%%%%%%%%%
\begin{figure}[t]
\centering
\includegraphics[width=8.4cm]{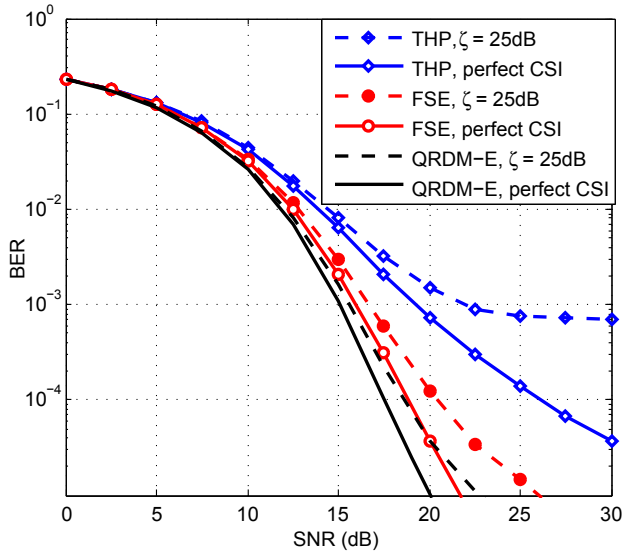}
\caption{Effect of the imperfect CSI on the BER performance of the QRDM-E, THP, and the proposed FSE-$p2$ algorithms in a $4 \times 4$ system, and using QPSK modulation.}
\label{fig:icsi}
\end{figure}
%%%%%%%%%%%%%%%%%%%%%%%%%%%%%
\subsection{\uppercase{Simulations with Imperfect CSI at the Transmitter}}
\label{sec:icsi}
In this section we consider that the channel state
information is not perfectly fed back to the transmitter due to
quantization error, practical channel estimation, feedback delay,
etc. Therefore, the channel matrix at the transmitter is defined as
\begin{equation}
\hat{\textbf{H}} = \textbf{H} + \textbf{B},
\end{equation}
where \textbf{H} is the perfectly estimated channel and \textbf{B} is the error matrix whose elements follow complex normal distributions with zero mean. Herein, we define $\zeta = 10\log_{10}\left(\left\|\textbf{H}\right\|^2/\left\|\textbf{B}\right\|^2\right)$ which is a measure of the amount of error.\\
\indent For sake of simplicity in the derivation, we consider that $\alpha \rightarrow 0$ and $N_t = N_u$, then the power of the error in the precoding matrix can be upper bounded by
\begin{equation}
e_{icsi} \leq \left(\sum_{i = 1}^{n} \frac{1}{\sigma_i(\textbf{H})^2}\right)^2 \left(\sum_{i = 1}^{n} \sigma_i(\textbf{B})^2\right),
\label{eq:eicsi}
\end{equation}
where $\sigma_i(\textbf{H})$ is the \textit{i}-th singular value of the \textbf{H} matrix. A detailed derivation of (\ref{eq:eicsi}) is given in the Appendix. It is clear from (\ref{eq:eicsi}) that a small error in the CSI may lead to a large error, particularly when the channel matrix is ill-conditioned where the first summation of (\ref{eq:eicsi}) becomes large.\\
\indent Figure \ref{fig:icsi} shows the BER performance of the precoding algorithms for $\zeta = 25$dB in a $4 \times 4$ system. At a target BER of $10^{-4}$, a degradation of 1 and 1.5dB are remarked in the BER performance curves of the QRDM-E and the proposed FSE-$p2$ vector perturbation techniques. This degradation is tolerable in practical systems compared to the remarkable reduction in the transmit power achieved by employing the proposed FSE. On the other hand, a floor appeared in the performance of the THP algorithm when $\zeta = 25$dB. This is because of the insufficiency of the number of candidates retained at each encoding level. Also, it indicates that the error in the precoding matrix becomes dominant.\\
\indent In \cite{QRDME} and \cite{Chae}, it has been shown that the
QRDME achieves the optimum diversity order, i.e., the BER
performance curve of the QRDME is parallel to that of the optimum
encoder. In this paper, we consider the QRDME as a reference, where
our proposed algorithm is shown to achieve the same diversity order
of the QRDME, i.e., they have parallel performance curves and
therefore our proposed algorithm achieves the optimum diversity.
%%%%%%%%%%%%%%%%%%%%%
\vspace{10pt}
\section{\uppercase{Conclusions}}
\label{sec:conc}
In this paper, we proposed a fixed-complexity sphere encoder (FSE) for MU-MIMO systems. Unlike the conventional SE scheme, which has a random complexity and a sequential structure, the proposed FSE has a fixed complexity and a parallel tree-search structure, leading to higher efficiency for hardware implementation. Moreover, the complexity of the FSE is analyzed and reduced by the two proposed techniques, viz.; pre-computing the frequently used values before the tree-search stage, and the comparison-before-squaring strategy which reduces the number of computations per node. Simulation and analytical results show that the proposed FSE requires a small fraction of the computational complexity and processing time of the conventional QRDM-E. This is achieved with a tolerable degradation in the BER while achieving the optimum diversity order.
\bibliographystyle{jcn}

%\bibliography{am_ger_eng,rubi_eng}

\appendix
Let the imperfect channel matrix be given by:
\begin{equation}
\hat{\textbf{H}} = \textbf{H} + \textbf{B},
\end{equation}
where \textbf{B} is the error matrix. Assume that the elements of \textbf{B} are small to assure that $\lim_{i \rightarrow \infty} \left(\textbf{H}^{-1}\textbf{B}\right)^i = \textbf{0}$, then, based on the \textit{Neumann series} we have:
\begin{align}
\left(\textbf{H} + \textbf{B}\right)^{-1} &= \left(\textbf{I} - [-\textbf{H}^{-1}\textbf{B}]\right)^{-1} \textbf{H}^{-1},\nonumber\\
                        &= \left(\sum_{i=0}^{\infty}[-\textbf{H}^{-1}\textbf{B}]^i\right) \textbf{H}^{-1}.
\label{al:Ney}
\end{align}
The first-order approximation of (\ref{al:Ney}) is given by:
\begin{equation}
\left(\textbf{H} + \textbf{B}\right)^{-1} \approx \textbf{H}^{-1} - \textbf{H}^{-1} \textbf{B} \textbf{H}^{-1}.
\end{equation}
Therefore, the approximated error in the precoding matrix becomes
$\left(\textbf{H}^{-1} - \hat{\textbf{H}}^{-1}\right) \approx
\textbf{H}^{-1} \textbf{B} \textbf{H}^{-1}$, and the
\textit{Frobenius norm} of this approximated error matrix is
upper-bounded as follows:
\begin{align}
\left\|\textbf{H}^{-1} \textbf{B} \textbf{H}^{-1}\right\|_{F}^{2} &\leq \left(\left\|\textbf{H}^{-1}\right\|_{F}^{2}\right)^2 \left\|\textbf{B}\right\|_{F}^{2},\nonumber\\
                    &= \left(\sum_{i = 1}^{n} \frac{1}{\sigma_i(\textbf{H})^2}\right)^2 \left(\sum_{i = 1}^{n} \sigma_i(\textbf{B})^2\right).
\label{eq:error}
\end{align}
\newpage
\epsfysize=3.2cm
\begin{biography}{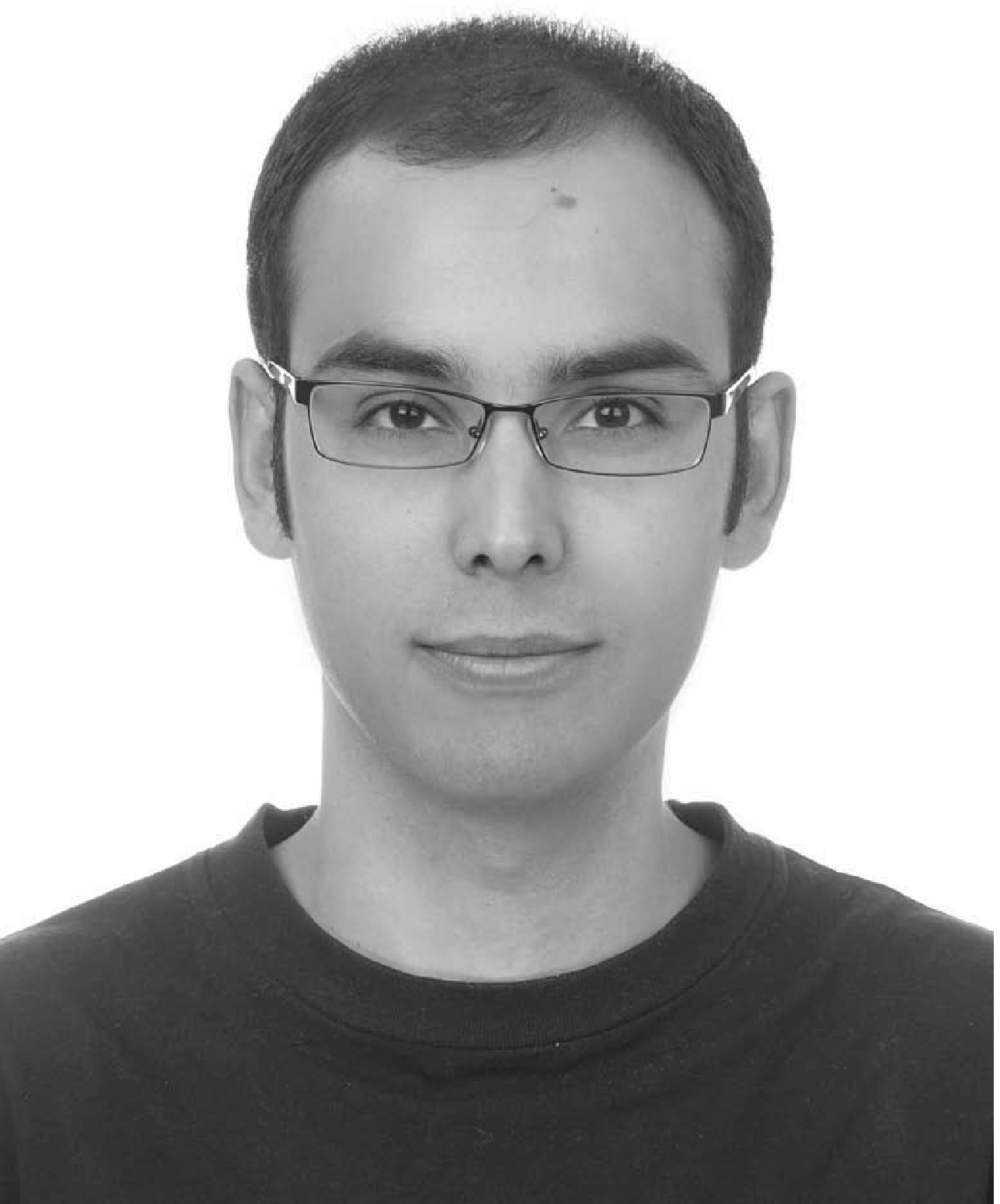}{Manar Mohaisen} received a B.Eng. in electrical engineering from the University of Gaza (IUG), Gaza, Palestine, in 2001. From 2001 to 2004, he was with the Palestinian Telecommunications Company - JAWWAL, Gaza, Palestine, where he worked as an operation and maintenance engineer and then as a cell-planning engineer. He received his M.S. degree in communication and signal processing from the University of Nice-Sophia Antiplois, Sophia Antipolis, France, in 2005. From March to September 2005, he followed an internship at IMRA Europe Co., Sophia Antipolis, France, as a part of his M.S. degree, where he worked on noise reduction in car environments. In February 2010, he obtained a Ph.D. degree from the Graduate School of Information Technology and Telecommunication, Inha University, Incheon, Korea. Since September 2010, he is with the School of Information Technology Engineering, Korea University of Technology and Education (KUT), where he is an assistant professor. His research interests include 3GPP LTE systems, detection schemes for spatial multiplexing MIMO systems, and precoding techniques for multiuser MIMO systems.
\end{biography}

\epsfysize=3.2cm
\begin{biography}{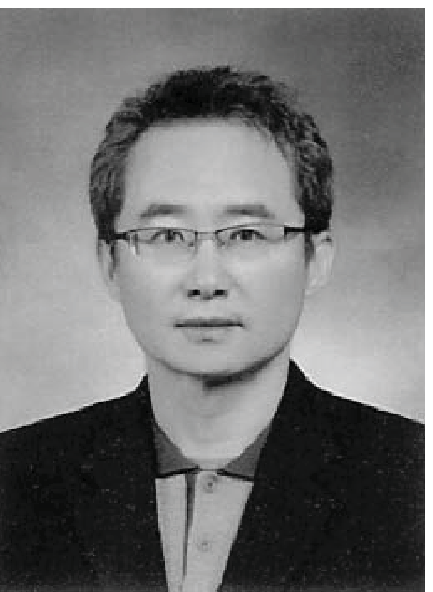}{KyungHi Chang} received his B.S. and M.S. degrees in electronics engineering from Yonsei University, Seoul, Korea, in 1985 and 1987, respectively. He received his Ph.D. degree in electrical engineering from Texas A$\&$M University, College Station, Texas, in 1992. From 1989 to 1990, he was with the Samsung Advanced Institute of Technology (SAIT) as a member of research staff and was involved in digital signal processing system design. From 1992 to 2003, he was with the Electronics and Telecommunications Research Institute (ETRI) as a principal member of technical staff. During this period, he led the design teams working on the WCDMA UE modem and 4G radio transmission technology (RTT). He is currently with the Graduate School of Information Technology and Telecommunications, Inha University, where he has been a professor since 2003. His current research interests include RTT design for IMT-Advanced system, 3GPP LTE and M-WiMAX system design, cognitive radio, cross-layer design, cooperative relaying systems, RFID/USN, and mobile Ad-hoc networks. Dr. Chang has served as a senior member of IEEE since 1998, and as an editor-in-chief of the Korean Institute of Communication Sciences (KICS) proceedings during 2007 - 2009. Currently, he is an editor-in-chief of the KICS Journal A. He has also served as an editor of ITU-R TG8/1 IMT.MOD, and he is currently an international IT standardization expert of the Telecommunications Technology Association (TTA). He is an recipient of the LG Academic Awards (2006), Haedong Best Paper Awards (2007), IEEE ComSoc Best Paper Awards (2008), and Haedong Academic Awards (2010).

\end{biography}

\end{document}